\documentclass{article}

\usepackage{amsmath,amsthm,amssymb}
\usepackage{graphicx}
\usepackage{fullpage}
\usepackage{url}
\usepackage{tikz}
\usetikzlibrary{calc}
\usepackage{xspace}

\DeclareMathOperator*{\E}{\mathbb{E}}
\let\Pr\relax
\DeclareMathOperator*{\Pr}{\mathbb{P}}

\newcommand{\A}{\texttt{A}}
\newcommand{\B}{\texttt{B}}
\newcommand{\C}{\texttt{C}}
\newcommand{\G}{\texttt{G}}
\renewcommand{\L}{\texttt{L}}

\newcommand{\U}{\texttt{U}}
\newcommand{\N}{\texttt{N}}

\newcommand{\M}{\texttt{B}}
\newcommand{\NI}{\texttt{NI}}
\newcommand{\HH}{\texttt{H}}

\renewcommand{\log}{\lg}

\newcommand{\NumActive}{\texttt{numActive}}

\newcommand{\Write}{\texttt{\textbf{write}}}
\newcommand{\Read}{\texttt{\textbf{read}}}
\newcommand{\Fill}{\texttt{\textbf{fill}}}
\newcommand{\Naive}{\texttt{naive}\xspace}

\newcommand{\prev}{\texttt{prev}}
\renewcommand{\next}{\texttt{next}}
\newcommand{\val}{\texttt{val}}

\newcommand{\True}{\texttt{True}\xspace}
\newcommand{\False}{\texttt{False}\xspace}
\newcommand{\LastFill}{$\Delta_{last}$\xspace}

\newcommand{\ceil}[1]{\left\lceil #1 \right\rceil}
\newcommand{\floor}[1]{\left\lfloor #1 \right\rfloor}

\newtheorem{theorem}{Theorem}
\newtheorem{definition}[theorem]{Definition}
\newtheorem{remark}[theorem]{Remark}

\author{
  Jacob Teo Por Loong\thanks{National University of Singapore High School of Math \& Science. \texttt{jacobtpl@gmail.com}. Work done while participating in the Research Science Institute, sponsored by the Center for Excellence in Education, in Summer 2017.}
  \and Jelani Nelson\thanks{Harvard University. \texttt{minilek@seas.harvard.edu}. Supported by NSF grant IIS-1447471 and CAREER award CCF-1350670, ONR Young Investigator award N00014-15-1-2388 and DORECG award N00014-17-1-2127, an Alfred P.\ Sloan Research Fellowship, and a Google Faculty Research Award.}
  \and Huacheng Yu\thanks{Harvard University. \texttt{yuhch@g.harvard.edu}. Supported in part by ONR grant N00014-15-1-2388, a Simons Investigator Award, and NSF grant CCF-1565641.}
}

\title{Fillable arrays with constant time operations and a single bit of redundancy }

\begin{document}

\maketitle

\begin{abstract}
In the {\em fillable array problem} one must maintain an array $\A[1..n]$ of $w$-bit entries subject to random access reads and writes, and also a $\Fill(\Delta)$ operation which sets every entry of $\A$ to some $\Delta\in\{0,\ldots,2^w-1\}$. We show that with just one bit of redundancy, i.e.\ a data structure using $nw+1$ bits of memory, \Read/\Fill\xspace can be implemented in worst case constant time, and $\Write$ can be implemented in {\em either} amortized constant time (deterministically) {\em or} worst case expected constant (randomized). In the latter case, we need to store an additional $O(\log n)$ random bits to specify a permutation drawn from an $1/n^2$-almost pairwise independent family.
\end{abstract}

\section{Introduction}\label{sec:intro}

A classic dynamic data structural problem is that of the {\em fillable array} \cite[Exercise 2.12]{AhoHU74}. In this problem, one wants to maintain an array $\A[1..n]$ with entries in $\{0,\ldots,2^w - 1\}$ subject to the following three operations:
\begin{itemize}
\item \Write$(i, \Delta)$\textbf{:} $\A[i] \leftarrow \Delta$
\item \Fill$(\Delta)$\textbf{:} $\A[i] \leftarrow \Delta$ for all $i=1..n$
\item \Read$(i)$\textbf{:} returns $\A[i]$
\end{itemize}
Note \Read$(i)$ may not be defined, if $\A[i]$ was never set due to a lack of a previous $\Fill$ or $\Write(i,\cdot)$ operation since the data structure's initialization. In this case, we allow the return value to be arbitrary (in fact, the data structures we present here return $0$ in this case, or some other pre-decided constant).

Most popular programming languages have some data structure implemented in its standard library supporting all these operations. For example, arrays in C/C++ can support \Fill\xspace via a call to \textbf{\texttt{memset}}, and a method even named $\Fill$ is implemented in C++ (for \texttt{ForwardIterator}), Python (\texttt{numpy.ndarray}), and Java (\texttt{Arrays}). In fact, arrays in Java must be filled with some value upon initialization as part of the language specification \cite{Java}.

The standard approach to implementing a fillable array uses $nw$ bits of memory, and in the word RAM model supports \Write/\Read\xspace each in $O(1)$ worst-case time and \Fill\xspace in time $O(n)$, simply via $n$ sequential writes. Recently \cite{HagerupK17} showed this is best possible for any data structure using $nw$ bits of memory. But what if we allow our data structure to use just a single bit of extra memory? Is is possible to then achieve all operations in worst case constant time? Despite the ubiquity of this problem, this basic question is unanswered.

For a data structure using $nw + r$ bits of memory, we denote the value of $r$ as the {\em redundancy}. The goal is to use as little redundancy as possible while supporting all three operations quickly. We assume the word RAM model with word size $w = \Omega(\log n)$, so that at the very least indexing into $\A\xspace$ can be performed in constant time.  A textbook exercise \cite[Exercise 2.12]{AhoHU74} shows that it is possible to achieve redundancy $r = 2n\ceil{\log_2 n} + \ceil{\log_2(n+1)} +  w$ bits while supporting all three operations mentioned above in worst case time $O(1)$. As in previous work, we refer to this data structure as the ``folklore'' solution. The same running time was achieved with better redundancy $r = (1+o(1))n$ by Navarro \cite{Navarro13}. Most recently, Hagerup and Kammer gave a solution with \Read/\Write\xspace time $O(t)$, \Fill\xspace time $O(1)$, and redundancy $r = \ceil{n/(w/(Ct))^t}$ for some constant $C>1$ for any desired integer $1\le t\le \log_2 n$ \cite{HagerupK17}. All these times are worst case. For $t = \log_2 n$, redundancy $r = 1$ is achieved.

\paragraph{Our main contribution.} We show it is possible to achieve $O(1)$ time for all three operations with redundancy $r=1$ if one settles for {\em amortized} complexity for \Write\xspace and worst-case complexity for \Read\xspace and \Fill. We also show that it is possible to replace the amortized $O(1)$ complexity for $\Write$ with $O(1)$ worst case {\em expected} running time, via a randomized data structure. In this case though, we need to store an additional $O(\log n)$ random bits to specify a permutation drawn from a $1/n^2$-almost pairwise independent family.

\bigskip

When describing our solutions, we assume $n$ is larger than some fixed constant since otherwise the trivial solution with $O(n)$ $\Fill$ time performs all operations in worst case time $O(1)$ with zero redundancy. We also henceforth use $[k]$ to denote $\{1,\ldots,k\}$ for integer $k$.

\section{Amortized solution}

We here describe and analyze our amortized solution, which is quite simple. The data structure operates in two {\em modes} and maintains a single mode bit which we refer to as \Naive. If \Naive is set to \True, then we are in {\em naive mode}. If set to \False, then we are in {\em linked list mode}. The single bit to store \Naive is the sole redundant bit in our representation, yielding $r=1$. This data structure, in either mode, also maintains an array $\B[1..n]$ such that each $\B[i]$ is a $w$-bit word. The data structure, when first initialized, starts in linked list mode.

We first describe naive mode. In this mode, we maintain the invariant that $\B[i] = \A[i]$ for all $i=1\ldots n$. Thus $\Write(i,\Delta)$ is implemented by performing the operation $\B[i]\leftarrow \Delta$, and $\Read(i)$ is executed by simply returning $\B[i]$. To execute $\Fill(\Delta)$, we set $\Naive$ to \False then initialize the data structure into linked list mode with value $\Delta$ (this initialization is to be explained shortly).

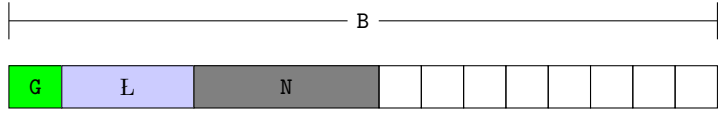
\begin{figure}
\begin{center}
\begin{tikzpicture}
	\foreach \i/\x in {0/0,1/20,2/70,3/140}
		\node [inner sep=0pt] at (\x pt, 0pt) (G\i) {};
	\foreach \i/\j/\l/\co in {0/1/\G/green,1/2/\L/blue!20,2/3/\N/black!50} {
		\draw [fill=\co] ($(G\i)+(0,8pt)$) rectangle ($(G\j)+(0,-8pt)$);
		\node at ($(G\i)!0.5!(G\j)$) {\small \l};
	}
	\foreach \x [evaluate=\x as \xx using {\x+16}] in {140,156,...,252}
		\draw (\x pt, 8pt) rectangle (\xx pt, -8pt);
	\node [inner sep=0pt] at (268pt, 0pt) (GE) {};
	\node at ($(G0)!0.5!(GE)+(0,25pt)$) (B) {\small \M};
	\draw ($(G0)+(0,25pt)$) -- (B) -- ($(GE)+(0,25pt)$);
	\draw ($(G0)+(0,18pt)$) -- ($(G0)+(0,32pt)$);
	\draw ($(GE)+(0,18pt)$) -- ($(GE)+(0,32pt)$);
\end{tikzpicture}
\caption{The organization of array $\B$ in linked list mode. $\B$ is divided into three subarrays, $\G, \L, \N$, and each cell is a $w$-bit word. The white cells in the array are unused in linked list mode, except during the process of conversion into naive mode triggered by $\NumActive$ reaching $n/C_L$ after a $\Write$.}\label{fig:ll-mode}
\end{center}
\end{figure}

Memory layout in linked list mode is depicted in Figure~\ref{fig:ll-mode}, together with the one extra \Naive bit not depicted there (set to \False). We say an index $i\in[n]$ is {\em active} if it has been written since the most recent initialization into linked list mode. $\G$ has size $2$ and stores the argument \LastFill to the last $\Fill$ call, as well as the number $\NumActive$ of active indices. $\L$ is an instance of the folklore data structure for an array with $\lceil\lg_2 n\rceil$ bit cells (sufficiently large to server as pointers into $\N$), and with array length $n/C_L$ for a constant $C_L > 1$ to be determined later. We abuse notation and let $\L[j]\leftarrow \Delta$ denote $\L.\Write(j, \Delta)$ and let $\L[j]$ denote the value returned by $\L.\Read(j)$. The main idea is that for each $j\in [n/C_L]$, $\L[j]$ is a pointer to the head node of a doubly linked list which contains all active indices $i$ in the range $\{(j-1)\cdot C_L+1,\ldots,j\cdot C_L\}$. For any such $i$, there is a node in the linked list containing the pair $(i, \A[i])$. Note that the linked list pointed to by $\L[j]$ is guaranteed to have at most $C_L = O(1)$ nodes. As mentioned in Section~\ref{sec:intro}, $\L$ occupies at most $3n/C_L + 2 \le 4n/C_L$ cells in $\B$. The actual linked list nodes are then allocated in the $\N$ array, which has a length that will be determined later. Each linked list node occupies $4$ $w$-bit cells, to store $\prev$ and $\next$ pointers (which are stored as indices into $\N$), as well as the two values $i$ and $\A[i]$ corresponding to that node. Null pointers are represented by the value $n$, which is unambiguous since $\N$ has size much less than $n$. The number of allocated nodes will always be equal to $\NumActive$, and thus whenever we wish to allocate a new node, we will do so by incrementing $\NumActive$ then using memory cells in the length-$4$ subarray $\N[(4\cdot(\NumActive-1) + 1)..(4\cdot\NumActive)]$.

Now we describe how to perform operations in linked list mode. To perform $\Fill(\Delta)$, no matter which mode we are in when the $\Fill$ was called, we set $\Naive$ to \False and do $\L.\Fill(null)$ (as mentioned previously, $null$ can be unambiguously represented by the value $n$ in this context). We also set $\NumActive$ to $0$ and \LastFill to $\Delta$. Initializing the entire data structure at the beginning of the operation sequence is identical, except that we set \LastFill to be $0$ (or whatever other pre-specified constant we would like to return when an $\A[i]$ value has never been set). Answering a $\Read(i)$ query is also simple. Set $j\leftarrow \floor{(i-1)/C_L}$. We first check whether $\L[j]$ is $null$. If so, we return \LastFill. Otherwise, we traverse the linked list $\L[j]$. If this list contains a node with a pair with index $i$, then we return the associated value in that node. Otherwise, we return \LastFill. Note $\Fill$ takes worst-case constant time as does $\Read$. This is because all $\Read/\Write/\Fill$ operations on $\L$ take constant time, and traversing $\L[j]$ during a $\Read$ takes time $O(C_L) = O(1)$.

The most involved operation to implement is the $\Write(i,\Delta)$ operation, which we now describe. We first determine whether $i$ was already active before this $\Write$ by performing the steps of $\Read(i)$. For $j = \floor{(i-1)/C_L}$ as defined above, note $i$ is active iff $\L[j]\neq null$ and the linked list $\L[j]$ contains a node with stored index $i$. If $i$ was already active, we simply ovewrite $\Delta$ as the associated value in the linked list node containing $i$. Otherwise, we increment $\NumActive$ then allocate a new node $v$ containing $(i,\Delta)$ and insert it to the front of the linked list $\L[j]$. If $\L[j]$ was $null$, then we set $\L[j]$ to the first cell of $v$ in $\N$. The main issue with this solution is that once $\NumActive$ is sufficiently large, we will run out of memory. This is because, on top the memory used to store $\G, \L$, every active index also uses up $4$ memory cells in $\N$. Since the number of active indices can be as big as $n$ and $\B$ only contains $n < 4n$ cells, we may run out of memory in $\N$ if the number of active indices becomes too large.

To avoid the above issue, we convert from linked list mode to naive mode whenever $\NumActive$ becomes too large; in particular, whenever it reaches $n/C_L$. Note then $\N$ need only be of length $4n/C_L$. To perform this conversion, we first set $\Naive\leftarrow \True$. We then set all white cells in $\B$ (see Figure~\ref{fig:ll-mode}) to $0$.  We then loop from $j = n/C_L$ down to $j^*$, for $j^*$ also to be determined later, and for each such $j$ we free all nodes in $\L[j]$. To free a node $v$ with prev/next pointers to $v.\prev$ and $v.\next$ and storing index $v.i$ and value $v.\val$, we first set $\B[v.i]\leftarrow \B[v.\val]$. We then set the $\next$ pointer of $\N[v.\prev]$ and $\prev$ pointer of $\N[v.\next]$ to point to each other, if not $null$. We then move the last node stored in $\N$ (which is stored in the $4$ cells starting at $4\cdot (\NumActive-1)+1$, inclusive) into the $4$ cells of $\N$ that used to store $v$. We then decrement $\NumActive$. In this way, during conversion into naive mode $\NumActive$ keeps track of the number of active indices that are yet to be converted into the naive representation. Note that if we divide $\A$ into contiguous blocks of length $C_L$, then active indices are converted into the naive representation in descending block order (though the order of conversion within a block may be arbitrary since linked lists are not sorted by index). We choose the value $j^*$ to be such that the $j^*$th block of indices in $\A$ is the closest block immediately to the right of the indices used in storing $\L$. In this way, the conversion continues until we pause midway, when we have converted all blocks of indices that do not intersect $\G,\L,\N$.

We now describe how to complete the conversion into naive mode, that is to convert all the indices in the remaining blocks $1,\ldots,j^* - 1$. Let the white part of the array $\B$ (see Figure~\ref{fig:ll-mode}) be denoted as subarray $\HH$. The idea here is to use {\em gaps} of three consecutive zeroes in $\HH$ to represent linked list nodes. Our goal is to build a linked list using the memory in these gaps to store all indices pointing to cells in $\G,\L,\N$ that are waiting to be converted. Let us now set some values. Note $\G,\L,\N$ combined use at most  $2 + 4n/C_L + 4 n/C_L = 8n/C_L + 2$ cells. As mentioned in Section~\ref{sec:intro}, we can assume $n$ is larger than some constant. In particular, we assume $n \ge 2C_L$ so that $8n/C_L + 2 \le 9n/C_L$. Thus we have $j^*-1 \le 9n/C_L^2 + 1 \le 10n/C_L^2$ assuming also $n\ge C_L^2/9$, and thus have at most $10n/C_L$ indices remaining to be converted. We need to make sure these cells can all be written into the gaps in $\HH$. Note $\HH$ has length at least $(1 - 9/C_L)n$ and contains a total of at most $n/C_L$ entries that are not zero (due to conversions of indices in blocks $j^*$ and above). Thus $\HH$ contains at least $\floor{(1 - 12/C_L)n / 3}$ disjoint gaps of three consecutive zero entries. We need $\floor{(1 - 12/C_L)n / 3} \ge 10n/C_L$ to ensure these items all fit in the gaps and $\HH$, and thus it suffices to set $C_L = 50$ for $n\ge 10$. Thus overall we have assumed $n \ge \max\{2 C_L, C_L^2/9, 10\} = 350$. We then use two pointers to simultaneously walk over the first $\NumActive$ nodes in $\N$ while walking over $\HH$, copying nodes into the gaps of three consecutive zeroes to form a link list in the gaps of $\HH$. We also use a single register during the conversion process to store the first cell of the first gap of three in $\HH$ (i.e.\ so that we know the head of the linked list). After we have finished copying over the remaining indices in $\N$ to the gaps in $\HH$, we then walk over the $\B$ entries used to store $\G,\L,\N$ then set them all to zero, then walk over the linked list in the gaps in $\HH$ and write the values of all these indices into their respective indices in index sections $\G,\L,\N$. We then perform one more walk over this gap linked list and rewrite zero in all its cells.

Note that this conversion process from linked list mode back to naive mode takes time $O(n)$, which can be charged to the $n/C_L$ active indices since the last $\Fill$. Thus overall this conversion process takes amortized time $O(1)$. 

\begin{theorem}
There is a deterministic data structure implementing fillable arrays with one bit of redundancy, supporting worst-case $O(1)$ time for $\Read/\Fill$ and $O(1)$ amortized time per $\Write$.
\end{theorem}

\section{Randomized solution}\label{sec:randomized}

In this section, we present a randomized implementation of a fillable array providing constant time per operation in expectation in the worst-case, and using one bit of redundancy. In fact, $\Read$ and $\Fill$ will take $O(1)$ time with probability $1$, whereas each $\Write$ will run in expected time $O(1)$. Our analysis assumes oracle access to a permutation $F$ drawn from an $1/n^2$-almost pairwise independent distribution of permutations on the set $[n]$. As we show in Appendix~\ref{sec:pairwise}, such an $F$ can be stored in $O(\log n)$ bits of space and evaluated in worst-case constant time on any $i\in[n]$, and it can be found in expected time $\mathop{poly}(\lg n)$ in pre-processing (see Remark~\ref{rem:primes}). We use the following standard definition of $\delta$-almost $k$-wise independent permutation families. See for example \cite{KaplanNR09}.

\begin{definition}
Let $D_1, D_2$ be distributions over a finite set $\Omega$. The {\em variation distance between} between $D_1$ and $D_2$ is
$$
\|D_1 - D_2\| := \frac 12\sum_{\omega\in \Omega}|D_1(\omega) − D_2(\omega)|
$$
We say that $D_1, D_2$ are $\delta$-close if $\|D_1 − D_2\|\le \delta$.
\end{definition}

\begin{definition}
Let $U_{\{n_k\}}$ denote the uniform distribution over the set of all $k$-tuples of distinct integers in $[n]$. A set $\Pi$ of permutations on $[n]$ is $\delta$-almost $k$-wise independent if for every $k$-tuple of distinct elements $x_1,\ldots,x_k\in[n]$, the distribution $(f(x_1), \ldots, f(x_k))$ for uniformly random $\pi\in \Pi$ is $\delta$-close to $U_{\{n_k\}}$.
\end{definition}

The high-level idea of the randomized solution is similar to the amortized solution presented in the previous section.
The data structure will have two modes: the \emph{naive mode} and the \emph{linked list mode}.
In the amortized solution, the only operation that takes more than constant time is when we need to convert the data structure from linked list mode to naive mode, which takes linear time.
However, this only happens after $\Theta(n)$ \Write{} operations after a \Fill.
To obtain expected worst-case constant time, the main idea is to gradually convert to naive mode over the $\Theta(n)$ \Write{} operations.
Since we put the last $C_L$ elements into the last linked list, it allows us to fill the last $C_L$ words of the array with their current values by going over the last linked list, and delete the last linked list.
Then we can view our data structure as in linked list mode for the first $n-C_L$ elements and in naive mode for the last $C_L$ elements.
However, if we keep inserting the elements that are in the first, say half, of the blocks, and convert to naive mode from the last blocks, we will at some point run out of space.
To avoid this issue, we apply a random permutation on the array \A, and prove that in expectation, we will ``run out of space'' only when there are a constant number of blocks left.
In the following, we present this approach with details.

\paragraph{The folklore solution with \texttt{delete}.}
The randomized solution we present in this section uses an implementation of the folklore solution supporting \texttt{delete} operation as a subroutine. 
More specifically, the subroutine maintains an array \A{} of length $n$ using $3n+2$ words, supporting
\begin{itemize}
	\item
		$\Read(i)$: return $\A[i]$;
	\item
		$\Write(i, \Delta)$: set $\A[i]$ to $\Delta$;
	\item
		$\Fill(\Delta)$: set $\A[i]$ to $\Delta$ for all $1\leq i\leq n$;
	\item
		$\texttt{delete}(n)$: deletes the last ($n$-th) element of the array \A, such that the data structure only uses the first $3(n-1)+2$ words of the memory. 
\end{itemize}

The subroutine supports every operation deterministically in constant time in worst case.
We defer the details to Appendix~\ref{sec:delete}.

\paragraph{Memory layout.}
As in the amortized solution, we refer to the one redundant bit as \Naive, which stores the mode of the data structure.
The rest of the data structure is stored in the memory \M{} of $n$ $w$-bit words. 

When \Naive is \True, the data structure is in naive mode.
In this case, we store $\A[i]$ in $\M[F(i)]$ for each $i$, where $F$ is the permutation previously mentioned.

\begin{figure}
\centering
\begin{tikzpicture}
	\foreach \i/\x in {0/0,1/20,2/70,3/140,4/220,5/260}
		\node [inner sep=0pt] at (\x pt, 0pt) (G\i) {};
	\foreach \i/\j/\l/\co in {0/1/\G/green,1/2/\L/blue!20,2/3/\N/black!50,3/4/\U/white,4/5/\NI/red} {
		\draw [fill=\co] ($(G\i)+(0,8pt)$) rectangle ($(G\j)+(0,-8pt)$);
		\node at ($(G\i)!0.5!(G\j)$) {\small \l};
	}
	\node at ($(G0)!0.5!(G5)+(0,25pt)$) (B) {\small \M};
	\draw ($(G0)+(0,25pt)$) -- (B) -- ($(G5)+(0,25pt)$);
	\draw ($(G0)+(0,18pt)$) -- ($(G0)+(0,32pt)$);
	\draw ($(G5)+(0,18pt)$) -- ($(G5)+(0,32pt)$);
\end{tikzpicture}
\caption{Memory layout of the data structure in linked list mode}\label{fig:mem_rand}
\end{figure}
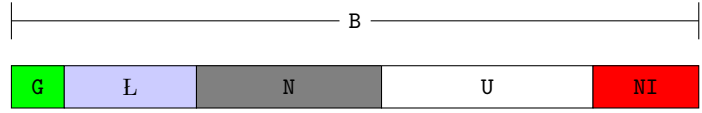

When \Naive is set to \False, the data structure is in linked list mode.
In this case, we partition the array \A{} into $\ceil{n/C_L}$ blocks.
The $j$-th block contains all the entries $i\in [n]$ such that $(j-1)\cdot C_L+1\leq F(i)\leq j\cdot C_L$.
Each block is associated with a \emph{doubly} linked list, in which, we store all elements that have been performed a \Write{} operation on since the last \Fill{}.
The $n$-word memory \M{} is partitioned into five subarrays in the following order (see Figure~\ref{fig:mem_rand}).
\begin{itemize}
	\item \G: this subarray has five words. The first four words store the pointers to the first word of the following subarrays. The last word stores \LastFill{}, the value to which the last \Fill{} operation sets.
	\item \L: this subarray stores a folklore data structure for the heads of all doubly linked lists.
	\item \N: this subarray stores all nodes in all linked lists. Each node has four fields, which are store in four words: the pointer to its predecessor, the pointer to its successor, the index and the value. To indicated the end of a linked list, the successor pointer of the last node will point to a word not in \N, e.g., the first word of \G. The same convention applies to \L{} when a linked list is empty, i.e., the header points to the first word of \G.
	\item \U: this subarray is unused.
	\item \NI: this subarray stores values of all entries that are mapped to this range by $F$, i.e., we set $\M[F(i)]=\A[i]$ for all $F(i)$ in this range.
\end{itemize}

\paragraph{Operation \Read$(i)$.}
If the data structure is in naive mode, the value of $\A[i]$ is stored in the $F(i)$-th word of the memory.
If the data structure is in linked list mode, we first check if $\M[F(i)]$ is in \NI{}, i.e., $\A[i]$ is converted to the naive mode already.
If it is, we simply return $\M[F(i)]$ as in the naive mode.
Otherwise, the $\floor{i/C_L}$-th block contains the element $\A[i]$.
We \Read{} the folklore data structure in subarray \L{}, and obtain the header of the linked list associated with this block.
Then we traverse this linked list to find all elements in the block that have been written since the last \Fill{}.
If $\A[i]$ is found in the linked list, we simply return its value stored in it.
Otherwise, we return \LastFill.

\paragraph{Operation \Write$(i, \Delta)$.}
If the data structure is in naive mode, we simply write the value $\Delta$ to the $F(i)$-th word.
If the data structure is in linked list mode, we first \Read{} the folklore data structure to find the header of the linked list associated with $\floor{i/C_L}$-th block.
Next we traverse the linked list to check if $\A[i]$ is already in it. 
If it is, we overwrite the value field of the node for $\A[i]$ with $\Delta$.
Otherwise, we allocate four more words at the end of \N, which can be done by increasing the pointer to subarray \U{} by four words.
Then we create a new node there for $\A[i]$, and insert it to the linked list.

An important idea of our randomized data structure is to gradually convert to naive mode.
Thus, in addition to the above procedure, we will perform \emph{convert} after each \Write{} operation.

\paragraph{Convert.}
The \emph{convert} procedure converts the last $C_T$ blocks into naive mode that are not converted yet, for some constant $C_T$ to be set later.
We first check if there is sufficient unused space (\U) left.
To do this, we first calculate the number of blocks $k$ that are still in linked list mode.
This number can be obtained from the size of \NI, i.e., $k=(n-|\NI|)/C_L$. 
If $|\U|\leq k\cdot C_L\cdot C_U$ or $k\leq 10$ for some constant $C_U$ (i.e., the data structure is running out of space soon), we run the linear time conversion algorithm on the remaining $k$ blocks as in the amortized solution, and set \Naive{} to \True.
Note that as long as we set $C_U$ to be greater than $0.95$, the linear time conversion algorithm will have sufficient working memory as we described in the previous section.

If $|\U|>k\cdot C_L\cdot C_U$ and $k>10$, we still have enough unused space, and can safely convert the blocks.
To convert the $k$-th block, we first decrease the pointer to \NI{} by $C_L$ words, and fill all of these $C_L$ words with \LastFill{}.
Next, we traverse the $k$-th linked list, and for all elements in the linked list, fill the $F(i)$-th word with the value of $\A[i]$. 
Then, we need to delete the last linked list. 
When deleting a node, we may create four unused words in the middle of \N. 
In this case, we simply move the last node in \N{} to this place, update the pointers and decrease the pointer to \U{} by four words. 
To delete the header, it suffices to run the \texttt{delete} operation on the folklore data structure (see Appendix~\ref{sec:delete}).
If a gap of more than four words is created between \L{} and \N, we again move the last node here, update the pointers and decrease the pointer to \U.

Finally, we repeat the above procedure $C_T$ times to convert the last $C_T$ blocks.

\paragraph{Operation \Fill$(\Delta)$.}
We set \Naive{} to \False{} no matter which mode the data structure was in, and set \LastFill{} to $\Delta$.
Then we update the pointers in \G{} such that \G{} has five words, \L{} has the size of a folklore solution on $\ceil{n/C_L}$ elements ($3\ceil{n/C_L}+2$), \N{} and \NI{} are empty, and \U{} has the remaining memory.
Finally we \Fill{} the folklore data structure with pointers to the first word of \G, i.e., empty all linked lists.

\paragraph{Analysis}
The correctness of the data structure is straightforward. 
It is also easy to verify that the only part of the data structure that may take super-constant time is the convert procedure. 

In the convert procedure, when too little unused space is left compared to the number of blocks remaining ($|\U|\leq k\cdot C_L\cdot C_U$), we convert all remaining blocks at once. 
In the following, we will show that this event happens with very small probability when the number of remaining blocks is large.

Fix a sequence of operations, and one operation in this sequence.
Now we analyze the expected time spent on this operation by the data structure.
If it is a \Fill{} or a \Read{}, the data structure does not invoke the convert procedure, and thus takes constant time in worst case. 
Otherwise, it is a \Write{} operation, and if the data structure has run a linear time conversion algorithm since the last \Fill{}, this \Write{} operation will take constant time in the worst case.
	
	Otherwise, let $k_U$ be the number of \Write{} operations since the last \Fill{}.
	The data structure invokes convert exactly once during each of the $k_U$ \Write s.
	The convert procedure converts $C_T$ blocks each time.
	Thus, we will have exactly $k=\ceil{n/C_L}-k_U\cdot C_T$ blocks left.

	Let $X$ be the number of entries written in those $k_U$ \Write{} operations and mapped to the first $k$ blocks, i.e., the number of elements that are inserted an still in linked list mode.
	We need to run a linear time conversion algorithm only when $|\U|\leq k\cdot C_L\cdot C_U$.
	On the other hand, we have 
	\[
	\begin{aligned}
		|\U|&\geq k\cdot C_L-|\N|-|\L|-|\G|-3\\
		&\geq k\cdot C_L-4X-(3k+2)-8 \\
		&= k\cdot C_L-4X-3k-10.
	\end{aligned}
	\]
	That is, we run the linear time conversion algorithm, only when
	\[
		X\geq \frac{1}{4}\left(k\cdot C_L\cdot (1-C_U)-3k-10\right).
	\]
	However, 
	\[
		\E X\leq k_U\cdot \frac{k\cdot C_L}{n}\leq \frac{n}{C_L\cdot C_T}\cdot \frac{k\cdot C_L}{n}=\frac{k}{C_T}, 
	\]
	which is much smaller.
	Now we are going to upper bound the probability using the pairwise independence of $F$.
	In fact, let $X_i$ be the indicator variable for the event that $F(i)\leq k\cdot C_L$, and let $S$ be the set of $k_U$ entries that are written after the last \Fill.
	By definition, we have the following:
	\begin{itemize}
		\item for each pair $X_i$, $X_j$, we have 
		\[
		\begin{aligned}
		\Pr(X_i=1\wedge X_j=1)&\leq \frac{k\cdot C_L}{n}\cdot \frac{k\cdot C_L-1}{n}+1/n^2 \\
		&= \Pr(X_i=1)\cdot \Pr(X_j=1)-k\cdot C_L/n^2+1/n^2 \\
		&\leq \Pr(X_i=1)\cdot \Pr(X_j=1)
		\end{aligned}
		\] by the pairwise almost independence;
		\item $X=\sum_{i\in S} X_i$;
		\item $\E X_i=\frac{k\cdot C_L}{n}$.
	\end{itemize}

	Thus, we set $C_L=100$, $C_T=8$ and $C_U=0.95$, and have
	\[
	\begin{aligned}
		&\color{white}\leq \color{black}\Pr\left(X\geq \frac{1}{4}\left(k\cdot C_L\cdot (1-C_U)-3k-10\right)\right) \\
		&\leq \Pr\left(X\geq \frac{1}{4}\cdot k\right) \\
		&\leq \Pr\left(\sum_{i\in S}(X_i-\E[X_i])\geq \frac{1}{8}\cdot k\right) \\
		&\leq\Pr\left(\left(\sum_{i\in S}(X_i-\E[X_i])\right)^2\geq \frac{1}{64}\cdot k^2\right) \\
		&\leq \frac{\E \left(\sum_{i\in S}(X_i-\E[X_i])\right)^2}{k^2/64}.
	\end{aligned}
	\]

	We also have
	\[
	\begin{aligned}
          &\color{white}=\color{black}\E\left(\sum_{i\in S}(X_i-\E[X_i])\right)^2 \\
		&=\sum_{i\in S} \E\left(X_i-\E[X_i]\right)^2 +2\sum_{i,j\in S,i<j} \E((X_i-\E[X_i])(X_j-\E[X_j]))\\
		&\leq k/8+2\sum_{i,j\in S,i<j} (\E(X_i-\E[X_i]))\cdot(\E(X_j-\E[X_j])) \\
		&\leq O(k).
	\end{aligned}
	\]

	Therefore, the probability that the data structure runs a $O(k)$-time conversion algorithm is at most $O(1/k)$, i.e., the expected running time on this operation is $O(1)$.\footnote{Note that we do not have to sum over all $k$, since each operation has a fixed $k$.}

\begin{theorem}
There is a Las Vegas randomized implementation of the fillable arrays with one bit of redundancy such that for any sequence of operations, each \Read{}/\Fill{} operation takes constant time in worst case, and each \Write{} operation takes constant time in expectation, assuming it has oracle access to a permutation $F$ drawn from a $1/n^2$-almost pairwise independent family of permutations over $[n]$.
\end{theorem}

As described in Section~\ref{sec:pairwise}, the permutation $F$ from an almost pairwise independent family can be represented in $O(\log n)$ bits of memory, sampled in $\mathop{poly}(\lg n)$ time, and evaluated in $O(1)$ time.

\section*{Acknowledgments}
We thank Omer Reingold explaining to us the construction in Appendix~\ref{sec:pairwise} (in fact a generalization of this approach that works for almost $k$-wise families for any $k$), and for allowing us to include a description of this construction.

\bibliographystyle{alpha}

\appendix

\section{Appendix}

\subsection{Almost pairwise independent permutations for all $n$}\label{sec:pairwise}

We here describe how to obtain an $O(1/n^c)$-almost pairwise independent permutation family $\Pi$ over $[n]$ of size $\mathop{poly}(n)$ for any $n$ larger than some constant, such that given a $O(\log n)$-bit description of some $\pi$ drawn randomly from $\Pi$ we can compute $\pi(i)$ for any $i$ in constant time. Here $c>0$ can be an arbitrary constant. This construction is used in Section~\ref{sec:randomized}.

We make use of the following theorem.

\begin{theorem}{\cite{Huxley72}}\label{thm:hux}
For any $\theta>7/12$ there exists a constant $n_0>0$ such that for all $n>n_0$, the interval $[n - n^\theta, n]$ contains $\Theta(n^\theta/\log n)$ prime numbers.
\end{theorem}

The starting point of the construction of $\Pi$ is the standard fact that for $n$ a prime, then $\Pi = \{ x\mapsto ax+b\mod n : a,b\in \mathbb F_n, a\neq 0\}$ is an exactly pairwise independent permutation family. We now describe how to extend this to an $O(1/n^{1-\theta})$-almost pairwise independent family over permutations on $[n]$ for arbitrary integer $n>n_0$ (for smaller $n$, one can just use the family of {\em all} permutations on $[n]$, which has constant size).

Pick a prime $p\in[n - n^\theta, n]$, which we know exists by Theorem~\ref{thm:hux}. Then $\pi\sim \Pi$ will be specified by picking three integers uniformly at random: $r\in\{0,\ldots,n-1\}$, and $a,b\in\mathbb F_p$ with $a\neq 0$. Define $\mathop{shift}_r(x) = x+r\mod n$. Then for $x\in[n]$ we define 
$$
\pi(x) = 
\begin{cases}
\mathop{shift}_r(x), & \text{ if } \mathop{shift}_r(x) \ge p\\
a\cdot \mathop{shift}_r(x) + b\mod p, & \text{ otherwise}
\end{cases}
$$
It is clear that any such $\pi$ is a permutation on $[n]$ and that $\pi(x)$ can be evaluated in worst case time $O(1)$, and furthermore a simple computation shows that $\Pi$ is $O(n^\theta/n)$-almost pairwise independent.

In order to decrease $\delta$ from $n^\theta/n$ down to $O(1/n^c)$, we use the following theorem of \cite{KaplanNR09}.

\begin{theorem}{\cite[Theorem 3.8]{KaplanNR09}}\label{thm:compose}
For a set of functions $\mathcal F$, let $\mathcal F^\ell$ denote the set of all functions $\{f_1\circ f_2 \cdots f_\ell : f_1,\ldots,f_\ell\in\mathcal F\}$ so that $|\mathcal F^\ell| = |\mathcal F|^\ell$. Then if $\Pi$ is a $\delta$-almost $k$-wise independent permutation family, then for any integer $\ell>1$, $\Pi^\ell$ is a $(\frac 12(2\delta)^\ell)$-almost $k$-wise independent permutation family.
\end{theorem}

Thus to decrease $\delta$, we can apply Theorem~\ref{thm:compose} with $\ell = \lceil c/(1-\theta)\rceil = O(1)$. The seed length and evaluation time to compute $\pi$ drawn randomly from $\Pi^\ell$ then both increase by only $O(1)$ factors.

\begin{remark}\label{rem:primes}
\textup{
Note that to apply the above construction, we need to find a prime $p\in[n-n^\theta, n]$ during pre-processing. By Theorem~\ref{thm:hux}, there are many such primes $p$ in this interval. In particular, we succeed in finding a prime with probability $\Omega(1/\lg n)$ by picking a random $p$ in this interval, which we can then test for primality in $\mathop{poly}(\lg n)$ deterministically \cite{AgrawalKS04}. Thus we can find this $p$ with a Las Vegas algorithm in pre-processing in expected time (and even with high probability) $\mathop{poly}(\lg n)$.
}
\end{remark}

\section{Folklore solution with delete}\label{sec:delete}

In this subsection, we present an implementation of the folklore data structure for fillable array $\A$ of length $n$ using $3n+2$ words of space.
Moreover, this implementation supports an extra operation \texttt{delete}$(n)$, which deletes the last ($n$-th) element in $\A$ such that the data structure only uses first $3(n-1)+2$ words of the memory after the operation.

The data structure will maintain the following variable/arrays:
\begin{itemize}
	\item \NumActive: the number of different elements written since the last \Fill
	\item \LastFill: the value that the last \Fill{} sets the array to
	\item \A: the array \A
	\item \B: first \NumActive{} entries store all elements written since the last \Fill
	\item \C: pointers to \B, i.e., $\B[\C[i]]=i$ if $i$ is written since the last \Fill
\end{itemize}

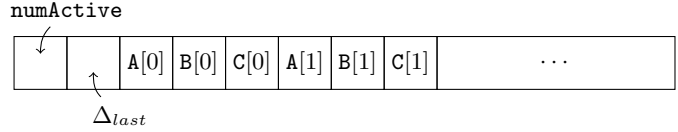
\begin{figure}\label{fig:folk}
\centering
\begin{tikzpicture}
	\foreach \i in {0,...,8}
		\node at ($(\i*20 pt, 0pt)$) (N\i) {};
	\node at (250pt, 0pt) (N9) {};
	\foreach \i/\j/\v in {0/1/,1/2/,2/3/\A[0],3/4/\B[0],4/5/\C[0],5/6/\A[1],6/7/\B[1],7/8/\C[1],8/9/\cdots} {
		\draw ($(N\i)+(0,10pt)$) rectangle ($(N\j)+(0,-10pt)$);
		\node at ($(N\i)!0.5!(N\j)$) (C\i) {\small $\v$};
	}
	\node at ($(C0)+(10pt,20pt)$) (CC0) {\small \NumActive};
	\node at ($(C1)+(10pt,-20pt)$) (CC1) {\small \LastFill};
	\draw [->] (CC0) to [in=90,out=225] (C0.90);
	\draw [->] (CC1) to [in=270,out=135] (C1.270);
\end{tikzpicture}
\caption{Memory layout of the folklore solution.}
\end{figure}

All three arrays \A, \B{} and \C{} has length $n$.
In total, the data structure uses $3n+2$ words of space.
To accomondate the \texttt{delete} operation, the three arrays will be interleaved with each other in memory (see Figure~\ref{fig:folk}).

Now we show how to implement the operations:
\begin{itemize}
\item To \Fill{} the array with $\Delta$, it suffices to set \NumActive{} to 0, and set \LastFill{} to $\Delta$.
\item To \Read{} $\A[i]$, we first check if $\C[i]\leq\NumActive$.
If $\C[i]>\NumActive$, we know $i$-th entry must have not been written since the last \Fill.
In this case, we return \LastFill.
If $\C[i]\leq\NumActive$, we then check if $\B[\C[i]]=i$.
If $\B[\C[i]]=i$, we return $\A[i]$.
Otherwise, we return \LastFill{}.
\item To \Write{} $\Delta$ to $\A[i]$, we first set $\A[i]$ to $\Delta$.
Next, we check if this is the first we write to $\A[i]$ since the last \Fill{} in the same way as we \Read{} $\A[i]$: check if $\C[i]>\NumActive$ or $\B[\C[i]]\neq i$.
If it is, we increment \NumActive{} by one, set \B[\NumActive] to $i$ and set $\C[i]$ to \NumActive.
\item To \texttt{delete} the last element, we first check if it has been written since the last \Fill.
If it has not, we do not have to do anything, and just ignore the last three words from now on.
Otherwise, we need to delete the record of $n$ in \B.
This can be done by moving \B[\NumActive] to $\B[\C[n]]$ and setting \C[\B[\NumActive]] to \C[n].
\end{itemize}

\end{document}